\documentclass{aa}

\usepackage[varg]{txfonts}
\usepackage[colorlinks=true, linkcolor=blue, citecolor=violet, urlcolor=gray]{hyperref}
\usepackage{xcolor}
\usepackage{listings}
\usepackage{amsmath}

\bibpunct{(}{)}{;}{a}{}{,}

\defcitealias{2012MNRAS.421.1007K}{K12}
\defcitealias{2014MNRAS.441.1340V}{V14}

\begin{document}

\title{The TNG50-SKIRT Atlas: \\
post-processing methodology and first data release}

\titlerunning{TNG50-SKIRT Atlas}

\author{%
Maarten Baes\inst{\ref{UGent}}
\and
Andrea Gebek\inst{\ref{UGent}}
\and
Ana Tr{\v{c}}ka\inst{\ref{UGent}}
\and
Peter Camps\inst{\ref{UGent}}
\and 
Arjen van der Wel\inst{\ref{UGent}}
\and
Abdurro'uf\inst{\ref{JHU},\ref{STScI}}
\and
Nick Andreadis\inst{\ref{UGent}}
\and
\\
Sena Bokona Tulu\inst{\ref{Jimma},\ref{UGent}}
\and
Abdissa Tassama Emana\inst{\ref{Jimma},\ref{UGent}}
\and
Jacopo Fritz\inst{\ref{UNAM}}
\and
Raymond Kelly\inst{\ref{BYU}}
\and
Inja Kova{\v{c}}i{\'{c}}\inst{\ref{UGent}}
\and
\\
Antonio La Marca\inst{\ref{SRON},\ref{Kapteyn}}
\and
Marco Martorano\inst{\ref{UGent}}
\and
Aleksandr Mosenkov\inst{\ref{BYU}}
\and
Angelos Nersesian\inst{\ref{UGent}}
\and
Vicente Rodriguez-Gomez\inst{\ref{UNAM}}
\and
\\
Crescenzo Tortora\inst{\ref{Napoli}}
\and
Bert Vander Meulen\inst{\ref{UGent}}
\and
Lingyu Wang\inst{\ref{SRON},\ref{Kapteyn}}
}

\institute{%
Sterrenkundig Observatorium, Universiteit Gent, Krijgslaan 281 S9, B-9000 Gent, Belgium\\ \email{maarten.baes@ugent.be}
\label{UGent}
\and
Department of Physics and Astronomy, N283 ESC, Brigham Young University, Provo, UT 84602, USA
\label{BYU}
\and
Center for Astrophysical Sciences, Department of Physics and Astronomy, The Johns Hopkins University, 3400 N Charles St., Baltimore, MD 21218, USA
\label{JHU}
\and
Space Telescope Science Institute, 3700 San Martin Drive, Baltimore, MD 21218, USA
\label{STScI}
\and
Physics Department, College of Natural Sciences, Jimma University, PO Box 378, Jimma, Ethiopia
\label{Jimma}
\and
Instituto de Radioastronom{\'{\i}}a y Astrof{\'{\i}}sica, Universidad Nacional Aut{\'{o}}noma de M{\'{e}}xico, Morelia, Michoac{\'{a}}n 58089, Mexico
\label{UNAM}
\and
SRON Netherlands Institute for Space Research, Landleven 12, 9747 AD, Groningen, The Netherlands
\label{SRON}
\and
Kapteyn Astronomical Institute, University of Groningen, Postbus 800, 9700 AV, Groningen, The Netherlands
\label{Kapteyn}
\and
INAF -- Osservatorio Astronomico di Capodimonte, Salita Moiariello 16, I-80131 Napoli, Italy
\label{Napoli}
}

\authorrunning{M. Baes et al.}

\date{\today}

\abstract{%
Galaxy morphology is a powerful diagnostic to assess the realism of cosmological hydrodynamical simulations. Determining the morphology of simulated galaxies requires the generation of synthetic images through 3D radiative transfer post-processing that properly accounts for different stellar populations and interstellar dust attenuation. We use the SKIRT code to generate the TNG50-SKIRT Atlas, a synthetic UV to near-infrared broadband image atlas for a complete stellar-mass selected sample of 1154 galaxies extracted from the TNG50 cosmological simulation at $z=0$. The images have a high spatial resolution (100~pc) and a wide field of view (160~kpc). In addition to the dust-obscured images, we also release dust-free images and physical parameter property maps with matching characteristics. As a sanity check and preview application we discuss the {\textit{UVJ}} diagram of the galaxy sample. We investigate the effect of dust attenuation on the {\textit{UVJ}} diagram and find that it affects both the star-forming and the quiescent galaxy populations. The quiescent galaxy region is polluted by younger and star-forming highly inclined galaxies, while dust attenuation induces a separation in inclination of the star-forming galaxy population, with low-inclination galaxies remaining at the blue side of the diagram and high-inclination galaxies systematically moving towards the red side. This image atlas can be used for a variety of other applications, including galaxy morphology studies and the investigation of local scaling relations. We publicly release the images and parameter maps, and we invite the community to use them.
}


\maketitle


\section{Introduction}

Cosmological hydrodynamical simulations are powerful tools to investigate the origin and evolution of galaxies. These simulations use numerical methods to emulate the behaviour of gas, stars, black holes, and dark matter in a virtual universe. They take into account the effects of gravity, hydrodynamics, star formation, feedback from supernovae and active galactic nuclei, and other physical processes that are known or expected to influence the formation and evolution of galaxies. For recent reviews, we refer to \citet{2015ARA&A..53...51S}, \citet{2020NatRP...2...42V}, and \citet{2023ARA&A..61..473C}.

In general, our confidence in the conclusions drawn from cosmological hydrodynamical simulations increases with the level of agreement between simulations and observations. Vice versa, detailed comparisons are required to calibrate the subgrid physics recipes in the simulations and to guide the improvements for the next-generation simulations. Only a decade ago, the overall agreement between hydrodynamical simulations and observations of galaxies was relatively poor \citep[e.g.][]{1991ApJ...380..320N, 1997ApJ...478...13N, 2000ApJ...538..477N, 1999ApJ...519..501S, 2012MNRAS.423.1726S}. In recent years the agreement  has improved significantly, mainly due to an increase in resolution and better implementations of subgrid physics recipes. Large-volume cosmological hydrodynamical simulations such as Illustris and IllustrisTNG \citep{2014MNRAS.444.1518V, 2018MNRAS.473.4077P}, EAGLE \citep{2015MNRAS.446..521S, 2015MNRAS.450.1937C}, Magneticum \citep{2016MNRAS.463.1797D}, Horizon-AGN and NewHorizon \citep{2016MNRAS.463.3948D, 2021A&A...651A.109D}, or SIMBA \citep{2019MNRAS.486.2827D} succeed in reproducing many observed global galaxy properties to a fair degree, including stellar and gas mass functions, the colour bimodality, the star-forming main sequence, the mass--metallicity relation, and many other scaling relations \citep[e.g.][to name just a few examples]{2017MNRAS.472.3354D, 2017MNRAS.467.2879B, 2019MNRAS.484.5587T, 2020MNRAS.497..146D, 2022ApJ...929...94A, 2022MNRAS.511.2544D, 2022MNRAS.512.6164Z}

Galaxies are no structureless point sources but complex ecosystems. The comparison of simulated galaxies to observations on local rather than global scales provides a powerful and meaningful test for cosmological hydrodynamical simulations. With many large imaging and integral-field spectroscopic surveys just finished, ongoing, or about to start, such as the SAMI Galaxy Survey \citep{2015MNRAS.447.2857B, 2021MNRAS.505..991C}, MaNGa \citep{2015ApJ...798....7B, 2022ApJS..259...35A}, the DESI Legacy Imaging Surveys \citep{2019AJ....157..168D}, WEAVE-StePS \citep{2023A&A...672A..87I}, or the Euclid Wide Survey \citep{2022A&A...662A.112E}, the required observational data are nowadays available. This offers the prospect of testing the simulation’s fidelity in much more detail, and to identify limitations due to both particle resolution or incorrect physics implemented in the models. 

Galaxy morphology is a particularly interesting characteristic to assess the realism of cosmological hydrodynamical simulations. Morphology has served as the main basis for galaxy classification since the early days of extragalactic astronomy \citep[][and references therein]{2005ARA&A..43..581S}. However, galaxy morphology is important beyond classification alone. It is a powerful diagnostic of the physical processes driving galaxy evolution (e.g.\ bars, mergers, spiral structure, tidal features). Morphological parameters are found to correlate with several fundamental galaxy properties such as stellar mass, colour, star formation history, merger history, and local galactic environment \citep{1980ApJ...236..351D, 2003ApJ...584..210G, 2003MNRAS.341...54K, 2003ApJS..147....1C, 2014ARA&A..52..291C, 2008ApJ...672..177L, 2008ApJ...675L..13V, 2009ARA&A..47..159B, 2014MNRAS.441..599B}.

In order to determine the morphology of simulated galaxies, we have to apply a forward modelling or post-processing approach on them. Since we know the 3D position and characteristics of all stellar particles in a cosmological hydrodynamical simulation, this seems a fairly straightforward projection operation. However, the generation of {\em{realistic}} synthetic images can be quite involved, especially due to the effects of absorption and scattering by dust grains in the interstellar medium. Indeed, dust attenuates a third to half of all the starlight in typical spiral galaxies in the local Universe \citep{2002MNRAS.335L..41P, 2016A&A...586A..13V, 2018A&A...620A.112B}, and detailed radiative calculations have shown that the effects of dust attenuation are often complex \citep{1992ApJ...393..611W, 1994ApJ...432..114B, 2004ApJ...617.1022P, 2006A&A...456..941M, 2010MNRAS.403.2053G}. The only way to generate synthetic images that properly account for dust absorption and scattering in a realistic geometry is by 3D radiative transfer modelling, a demanding and challenging numerical problem \citep{2013ARA&A..51...63S}.

In the past few years, there have been several efforts to generate synthetic images for large sets of galaxies extracted from cosmological hydrodynamical simulations. \citet{2015MNRAS.447.2753T} used the SUNRISE radiative transfer code \citep{2006MNRAS.372....2J} to generate a broadband image atlas for about 7000 $z=0$ galaxies extracted from the Illustris simulation (without including dust in their models). \citet{2015MNRAS.452.2879T} used the SKIRT code \citep{2015A&C.....9...20C, 2020A&C....3100381C} to produce an image atlas for more than 30,000 galaxies from the EAGLE simulation at $z = 0.1$, fully accounting for dust attenuation. Similarly, \citet{2019MNRAS.483.4140R} generated realistic synthetic Pan-STARRS-like images for more than 12,000 galaxies from the TNG100 simulation using SKIRT. The power of these large-volume-simulation-based image databases is that they cover the entire galaxy population, from massive ellipticals to less massive star-forming galaxies. The drawback is that the physical resolution of the underlying simulations, and hence also of the mock images, is limited to about 1~kpc, a limiting factor in both realism and range of applications. 

This disadvantage can be addressed by shifting from large-volume to higher-resolution zoom-in simulations, such as FIRE \citep{2014MNRAS.445..581H}, NIHAO \citep{2015MNRAS.454...83W}, APOSTLE \citep{2016MNRAS.457.1931S}, Latte \citep{2016ApJ...827L..23W}, Auriga \citep{2017MNRAS.467..179G}, RomulusC \citep{2019MNRAS.483.3336T}, or ARTEMIS \citep{2020MNRAS.498.1765F}. \citet{2021MNRAS.506.5703K} and \citet{2022MNRAS.512.2728C} generated a suite of synthetic high-resolution images for the present-day galaxies from the Auriga and ARTEMIS simulations, respectively, with a combined image database that covers the entire UV--submm wavelength range. \citet{2023ApJ...957....7F} also used SKIRT to generate UV--submm images for a set of present-day galaxies from the NIHAO simulation.  The main limitation of these studies is that they only cover a relatively small number of galaxies (30 from Auriga, 45 from ARTEMIS, 65 from NIHAO) and the limited dynamical range in galaxy properties.

The TNG50 simulation \citep{2019MNRAS.490.3196P, 2019MNRAS.490.3234N} offers the prospect of combining the advantages of both approaches as it combines a high spatial and mass resolution typical for zoom-in simulations with a large number and diversity of galaxies typical for large-volume simulations. Very recently, \citet{2023MNRAS.519.4920G} presented a database of simulated KiDS-like {\em{gri}} images for more than 5000 galaxies from the TNG50 simulation at $z=0.034$ to investigate the connection between galaxy mergers and optical morphology in the local Universe. Their approach is similar to \citet{2019MNRAS.483.4140R} and fully accounts for dust attenuation using SKIRT. At the same time, several teams have generated synthetic optical data cubes for samples of galaxies extracted from the TNG50 simulation, with different levels of refinement in the way dust effects are approximated \citep{2022MNRAS.514.2821B, 2022MNRAS.515..320N, 2023MNRAS.522.5479N, 2023A&A...673A..23S}.

In this paper we present and release the TNG50-SKIRT Atlas (hereafter TSA), a synthetic image database for 1154 galaxies at $z=0$ extracted from the TNG50 simulation. We cover a wide range of galaxy properties, we generate mock images in 18 broadband filters in the UV to near-infrared (NIR) wavelength range with high spatial resolution and a wide field of view, while fully accounting for dust absorption and scattering. Moreover, we also generate images that are free of dust attenuation, as well as synthetic maps of intrinsic physical parameters such that connections between the observed and the intrinsic properties can easily be made.

This paper is organised as follows. In Sect.~{\ref{TNG50.sec}} we briefly present the TNG50 cosmological hydrodynamical simulation and the SKIRT radiative transfer code, and we discuss our methodology to generate synthetic images for the TNG50 galaxies. In Sect.~{\ref{ImageAtlas.sec}} we present the multi-wavelength image database. In Sect.~{\ref{UVJ.sec}} we discuss the {\textit{UVJ}} diagram of the galaxy sample as an illustration of the range of applications that is possible with these data. We present a summary and our conclusions in Sect.~{\ref{Conclusions.sec}}.


\section{Synthetic images for the TNG50 simulation}
\label{TNG50.sec}

\subsection{The TNG50 simulation}

The TNG50 simulation \citep{2019MNRAS.490.3196P, 2019MNRAS.490.3234N} is the highest-resolution version of the IllustrisTNG cosmological magneto-hydrodynamical simulations suite \citep{2018MNRAS.480.5113M, 2018MNRAS.477.1206N, 2018MNRAS.475..624N, 2018MNRAS.475..648P, 2018MNRAS.475..676S}. It follows the evolution of a cubic volume of 51.7 comoving Mpc on the side, with cosmological parameters based on the Planck 2015 results, namely $\Omega_{\text{m}}= 0.3089$, $\Omega_{\text{b}}= 0.0486$, $\Omega_\Lambda = 0.6911$, and $H_0 = 67.74~{\text{km}}\,{\text{s}}^{-1}\,{\text{Mpc}}^{-1}$ \citep{2016A&A...594A..13P}. TNG50 uses the moving-mesh code AREPO \citep{2010MNRAS.401..791S} as its hydrodynamics solver. The galaxy formation model \citep{2017MNRAS.465.3291W, 2018MNRAS.473.4077P} in the TNG50 simulation is identical for all three TNG simulations. The physical processes accounted for in the simulation include gas cooling and heating, stochastic star formation, stellar evolution, chemical enrichment of the ISM, feedback from supernovae, seeding and growth of supermassive black holes, AGN feedback, and magnetic fields. TNG50 reaches a baryonic mass resolution of $8.5\times10^4~{\text{M}}_\odot$; the average cell size in the star-forming regions of galaxies is 70--140~pc. For a full description of the TNG50 simulation we refer to \citet{2019MNRAS.490.3196P} and \citet{2019MNRAS.490.3234N}.

\subsection{Sample selection}

\begin{figure}
\includegraphics[width=\columnwidth]{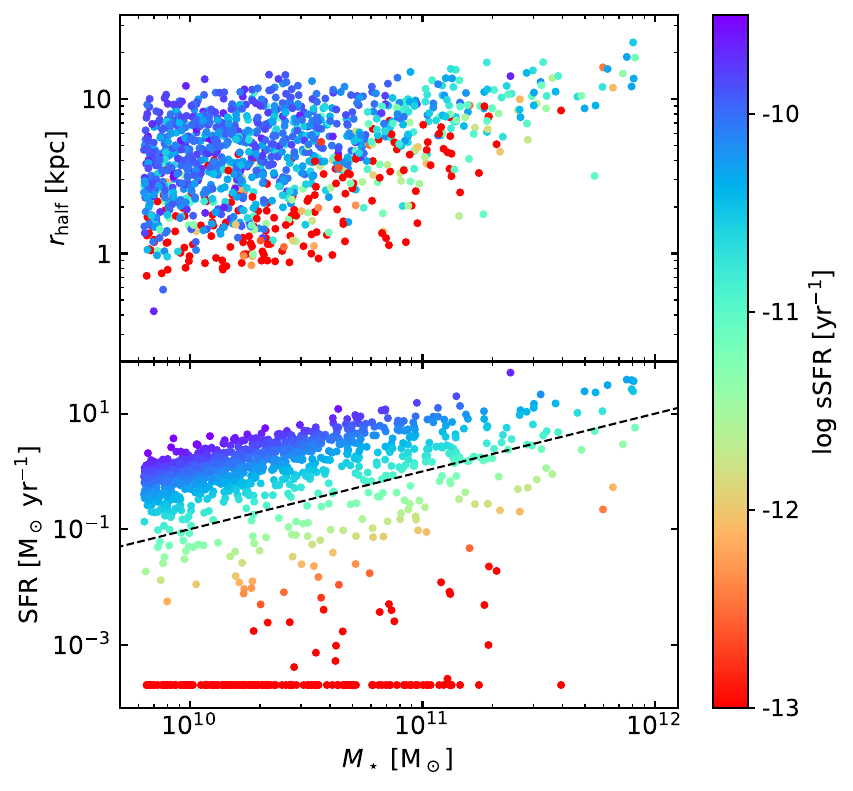}
\caption{Relation between stellar mass, half-mass radius, and SFR for the TNG50 galaxies in our sample. In the bottom panel, the dashed line indicates the separation between quiescent and star-forming galaxies (${\text{sSFR}} = 10^{-11}~{\text{yr}}^{-1}$). Galaxies with no ongoing star-formation are plotted at ${\text{SFR}} = 2\times10^{-4}~{\text{M}}_\odot~{\text{yr}}^{-1}$.}
\label{sample.fig}
\end{figure}

To build the TSA, we selected all TNG50 galaxies at $z=0$ with total stellar mass between $10^{9.8}$ and $10^{12}~{\text{M}}_\odot$ from the TNG50 public database \citep{2019ComAC...6....2N}, resulting in a sample of 1154 individual galaxies. The characteristics of the sample are illustrated in Fig.~{\ref{sample.fig}}, which shows the correlations between stellar mass, star formation rate (SFR), and half-mass radius. As evident from the bottom panel, most galaxies in the sample are star-forming galaxies lying in the main sequence, whereas there is also a population of more quiescent galaxies. Adopting a specific star formation rate (sSFR) of ${\text{sSFR}} = 10^{-11}~{\text{yr}}^{-1}$ as the boundary between star-forming and quiescent galaxies \citep[e.g.][]{2004MNRAS.351.1151B, 2009MNRAS.397.1776F, 2019MNRAS.485.4817D, 2023A&A...669A..11P}, our sample contains 869 star-forming and 285 quiescent galaxies.

\subsection{The SKIRT radiative transfer code}

SKIRT \citep{2015A&C.....9...20C, 2020A&C....3100381C} is a three-dimensional Monte Carlo radiative transfer code. Originally set up as a dust radiative transfer code and designed to model the effects of dust in galaxies \citep{2003MNRAS.343.1081B, 2011ApJS..196...22B}, it has transformed into a more generic Monte Carlo radiative transfer tool. The physical ingredients within SKIRT include absorption, multiple scattering, and stochastic emission \citep{2015A&A...580A..87C}; polarisation caused by scattering or by emission from aligned nonspherical dust grains \citep{2017A&A...601A..92P, 2021A&A...653A..34V}; Ly$\alpha$ resonant line scattering \citep{2021ApJ...916...39C}; absorption and emission at rotational or electronic transitions for selected ions, atoms and molecules \citep{2023MNRAS.521.5645G, 2023A&A...678A.175M}; photo-absorption, fluorescence, and scattering at X-ray wavelengths \citep{2023A&A...674A.123V}. SKIRT is equipped with a library of flexible input models, routines to import the output from various kinds of hydrodynamical simulations, and a selection of advanced spatial grids for discretising the medium \citep{2015A&C....12...33B, 2013A&A...554A..10S, 2014A&A...561A..77S, 2013A&A...560A..35C}. Many Monte Carlo radiative transfer optimisation mechanisms and a hybrid parallelisation strategy are implemented to maximise the efficiency of the code \citep{2008MNRAS.391..617B, 2013ARA&A..51...63S, 2016A&A...590A..55B, 2022A&A...666A.101B, 2017A&C....20...16V}.

While there are several applications in other fields \citep[e.g.][]{2015A&A...577A..55D, 2016MNRAS.460.3975H, 2021MNRAS.507.5246M, 2022A&A...662A..81E, 2023ApJ...950...88J}, SKIRT is mainly used in an extragalactic context. It is used to generate images, spectra, spectral energy distributions, and polarisation maps for idealised galaxy models \citep{2010MNRAS.403.2053G, 2014MNRAS.441..869D, 2016MNRAS.463.2912L, 2021MNRAS.507.2755L, 2022ApJ...930...66Q}, for high-resolution models fitted to observed spiral galaxies \citep{2014A&A...571A..69D, 2017A&A...599A..64V, 2020A&A...638A.150V, 2018A&A...616A.120M, 2019MNRAS.487.2753W, 2020A&A...637A..24V, 2020A&A...637A..25N, 2020A&A...643A..90N}, and for simulated galaxies extracted from cosmological simulations \citep[e.g.][]{2015A&A...576A..31S, 2017MNRAS.470..771T, 2019MNRAS.483.4140R, 2020MNRAS.492.5167V, 2021MNRAS.502.3210L, 2023MNRAS.519.2475H, 2023MNRAS.518.5522C}.

\subsection{The SKIRT setup to generate high-resolution images}
\label{SKIRT_setup.sec}

Each SKIRT simulation is completely determined by specifying the characteristics of the primary radiation sources, the transfer medium, a suite of instruments to capture the emerging radiation field, probes to measure other aspects of the simulation, and a number of technical or numerical simulation parameters. The recipe for the SKIRT simulations we adopted in this work largely follows the prescriptions outlined by \citet{2022MNRAS.516.3728T}, which were inspired by earlier works by \citet{2016MNRAS.462.1057C, 2018ApJS..234...20C, 2022MNRAS.512.2728C} and \citet{2021MNRAS.506.5703K}. We only give a brief overview, focusing on the particular aspects of the current work, and refer to these papers for more details. 

\subsubsection{Primary radiation sources}

For each galaxy, the primary radiation sources are the stellar particles that belong to the galaxy, which were extracted from the TNG50 database. To each stellar particle older than 10 Myr, we assigned a simple stellar population (SSP) SED from the \citet{2003MNRAS.344.1000B} SSP family with a \citet{2003PASP..115..763C} initial mass function. The mass, metallicity, and age of the SSP are directly inherited from the TNG50 particle data. Each particle was given a smoothing length corresponding to the distance to the 32nd nearest neighbour, with a maximum smoothing length of 800~pc. 

Stellar particles younger than 10 Myr are assumed to be still partly embedded in their birth cloud. They were assigned an SED from a library of H{\sc{ii}} region templates based on the MAPPINGS~III library \citep{2008ApJS..176..438G}. Each template in this library is characterised by five free parameters, two of which (metallicity and SFR) can be directly obtained from the particle data. Since the specific choice of the ISM pressure value hardly affects the broadband SED shape \citep[][Fig. 4]{2008ApJS..176..438G}, we used a fixed value, $P = 1.38\times10^{-12}~{\text{Pa}}$. For the compactness, which essentially determines the dust temperature and thus the shape of the far-infrared spectrum, we sampled a value from a lognormal distribution \citep{2021MNRAS.506.5703K, 2022MNRAS.516.3728T} with parameters calibrated on the dust temperature distribution in observed and simulated star forming regions \citep{2019ApJ...874..141U, 2020MNRAS.499.5732K}. Finally, the PDR covering factor was set to $f_{\text{PDF}} = {\text{e}}^{-t/\tau}$ with $t$ the particle age and $\tau$ the PDR clearing timescale, a free parameter in the post-processing recipe. The value $\tau = 3~{\text{Myr}}$ was determined by \citet{2022MNRAS.516.3728T} from a calibration of the integrated TNG50 fluxes against observational data from the DustPedia nearby galaxy sample \citep{2017PASP..129d4102D, 2018A&A...609A..37C}.

\subsubsection{The dusty medium}

The properties of the transfer medium, in our case the dusty interstellar medium, are based on the characteristics of the Voronoi gas cells in the hydrodynamical simulation, which were again downloaded from the TNG50 database. The density of the dust at every location in the SKIRT simulation volume is based on the assumption that a fixed fraction $f_{\text{dust}}$ of the metals in the ISM gas is locked up in dust grains, that is
\begin{equation}
\rho_{\text{dust}} = 
\begin{cases}
\;f_{\text{dust}}\,Z_{\text{gas}}\,\rho_{\text{gas}}
&\quad{\text{if ISM}},\\
\;0
&\quad{\text{else}},
\end{cases}
\end{equation}
with $Z_{\text{gas}}$ the metallicity and $\rho_{\text{gas}}$ the density of the gas. To determine which gas cells qualify as ISM gas cells, we applied the prescription by \citet{2012MNRAS.427.2224T, 2019MNRAS.484.5587T} that separates the hot circumgalactic medium from the cooler ISM gas. In this framework, gas is considered as belonging to the ISM if its temperature $T_{\text{gas}}$ satisfies the condition
\begin{equation}
\log\left(\frac{T_{\text{gas}}}{\text{K}}\right) < 6 + 0.25 \log\left(\frac{\rho_{\text{gas}}}{10^{10}~h^2~{\text{kpc}}^{-3}}\right).
\end{equation}
The density, metallicity and temperature of each gas cell were directly taken from the cell data. The only free parameter left was the dust-to-metal fraction $f_{\text{dust}}$, for which \citet{2022MNRAS.516.3728T} determined $f_{\text{dust}} = 0.2$ as the best value in combination with the PDR clearing timescale of $\tau = 3~{\text{Myr}}$ (see previous section). This value is comparable to observational estimates based on different nearby galaxy samples \citep[e.g.][]{2019A&A...623A...5D, 2021A&A...649A..18G, 2021MNRAS.502.4723Z}.

We assumed a fixed dust grain model at every location in the galaxy. As our dust grain model, we used the diffuse ISM THEMIS model \citep{2017A&A...602A..46J}. It consists of a distribution of carbonaceous and silicate grains, with optical properties based on laboratory data where possible, and can reproduce many observed properties of the dust in the Milky Way, including the UV to infrared extinction curve, extinction correlations, the thermal dust emission spectrum, and red and blue luminescence. For the present paper we focused on the UV to near-infrared wavelength range, and we did not include diffuse dust emission in the radiative transfer procedure (ISM dust emission is only important in the mid-infrared to mm spectral range). Such extinction-only SKIRT simulations are much faster than simulations that take dust emission into account, such as the work by \citet{2021MNRAS.506.5703K} and \citet{2022MNRAS.512.2728C}.

The SKIRT code requires a grid structure on which the medium is discretised. Since the TNG50 simulation uses the AREPO moving mesh technique to solve the hydrodynamics, we could have opted to use the native Voronoi grid for the SKIRT post-processing as well. SKIRT is equipped with a mechanism to efficiently traverse photon packets through a Voronoi grid \citep{2013A&A...560A..35C}, and this approach has been adopted to post-process moving-mesh hydrodynamical simulations \citep[e.g.][]{2019MNRAS.483.4140R, 2020MNRAS.497.4773S, 2021ApJ...916...39C, 2022MNRAS.510.3321P, 2023MNRAS.519.4920G}. However, like in many other works \citep[e.g.][]{2020MNRAS.492.5167V, 2021MNRAS.506.5703K, 2022MNRAS.516.3728T, 2023MNRAS.519.2475H, 2023MNRAS.524..907B} we chose to re-grid the dust density distribution onto a hierarchical octree grid to boost the speed of the radiative transfer process. We used up to 12 levels of refinement to ensure that we fully resolve the dust density distribution.

\subsubsection{Instruments and probes}

\begin{table}
\caption{Polar angle ($\theta$) and azimuth ($\phi$) of the observer positions with respect to the simulation volume.}
\label{Obs.tab}
\centering
\begin{tabular}{crr}
observer & $\theta$ [deg]\hspace*{1em} & $\phi$ [deg]\hspace*{1em} \\[0.5em] \hline \\[-0.2em]
O1 & $60.279459$  & $99.800435$ \\
O2 & $98.997076$ & $-19.706009$ \\
O3 & $109.839470$ & $-146.239796$ \\
O4 & $30.457794$ & $-94.086488$ \\
O5 & $149.542206$ & $85.913512$ \\[0.5em] \hline
\end{tabular}
\end{table}

\begin{table*}
\caption{Main characteristics of the TNG50-SKIRT Atlas.}
\label{AtlasChar.tab}
\centering
\begin{tabular}{ll}
\hline \\[-0.2em]
parent simulation & TNG50-1 (snapshot 99, $z=0$) \\
number of galaxies & 1154 \\
stellar mass range & $10^{9.8} - 10^{12}~{\text{M}}_\odot$ \\ 
observer positions & 5 random positions (see Table~{\ref{Obs.tab}}) \\
broadband filters & 18 filters (GALEX, Johnson {\textit{UBVRI}}, LSST, 2MASS, WISE W1 and W2) \\
dust attenuation treatment & dust-obscured and dust-free images \\
physical parameter maps & $\Sigma_\star$, $\Sigma_{\text{dust}}$, $\langle Z_\star \rangle$, $\langle t_\star \rangle$ \\
spatial resolution of images/maps & 100~pc \\
number of pixels & $1600 \times 1600$ \\
field-of-view & 160~kpc \\
images/maps per galaxy & $5 \times 40 = 200$ \\ 
total number of images/maps & $1154 \times 200 = 230\,800$ \\
total data volume & 1.97 Tb \\
data access location & \url{https://skirt.ugent.be/data/tng50-skirt-atlas} \\[0.5em] \hline
\end{tabular}
\end{table*}

We defined a set of broadband imagers as instruments in our SKIRT simulation. Contrary to \citet{2021MNRAS.506.5703K} and \citet{2022MNRAS.512.2728C}, who positioned the observers at specific sight lines relative to orientation of each galaxy (face-on, edge-on, and at specific inclination angles), we aimed at arbitrary viewing points. We used fixed viewing points relative to the simulation box, which has no connection to the orientation of each individual galaxy. In order to limit the computation time and the data volume, we settled on $N_{\text{obs}} = 5$ viewing positions per galaxy. These positions were spread on the unit sphere in optimal arrangement, that is, so as to maximise the angular distance between them. This problem is generally known in geometry as the Tammes problem \citep{Tammes1930}. Exact solutions for low values of $N_{\text{obs}}$ are available in the literature \citep[e.g.][]{FejesToth1943, Schutte1953, Erber1991}, as well as approximate numerical solutions for larger values of $N_{\text{obs}}$ \citep[e.g.][]{Kottwitz1991, Hardin1994, 1996JQSRT..56...97S}. For the case $N_{\text{obs}} = 5$, two of the observer positions are antipodal, and this semi-redundancy provides an opportunity to test the accuracy of any analysis method on the images. The details of the observer positions are provided in Table~{\ref{Obs.tab}}.

For the instruments, a pixel scale of 100 pc provides a nice match to the spatial resolution of the TNG50 simulation \citep{2019MNRAS.490.3196P, 2019MNRAS.490.3234N}. The actual size of the detectors, or equivalently, the field-of-view, does not significantly affect the simulation run time, but it does impact the data volume. We chose detectors with $1600\times1600$~pixels, corresponding to a field of view of 160~kpc on the side, in order to cover the outer regions of the most extended galaxies. 

SKIRT offers the opportunity to define broadband data cubes, that is, synthetic instruments that contain multiple broadband images automatically convolved with the correct transmission curves \citep[][Sect.~4.5]{2020A&C....3100381C}. We generated images in the GALEX FUV- and NUV-bands, the Johnson {\it{UBVRI}}-bands, the LSST {\it{ugrizy}}-bands, the 2MASS {\it{JHK}}$_{\text{s}}$-bands, and the WISE W1- and W2-bands. Together, these bands cover the wavelength range between 0.1 to 5~$\mu$m, the domain of dominance of stellar emission. 

Apart from instruments that record the radiation escaping from the simulation volume, SKIRT offers the opportunity to install probes, which sample numerical or physical quantities internal to the simulated model. Probes thus provide relevant diagnostics on the simulation setup and offer an opportunity to investigate properties of the simulated model that could never be directly observed from the outside, such as for example the radiation field \citep[][Sect.~3.5]{2020A&C....3100381C}. Specifically for this project, a set of probes was implemented in SKIRT to generate intrinsic physical parameter maps for an arbitrary observer's position. These maps are generated by projecting the 3D physical fields such as the stellar mass density on the observer's plane of the sky. Concretely, we generated stellar mass surface density maps, stellar-mass-weighted metallicity and age maps, and dust mass surface density maps for each of the five observers, with the same pixel scale and field-of-view as the instruments described above.

In a similar way, the emissivity of the stellar particles within a given broadband can be projected on an observer's plane of the sky, resulting in dust-free broadband images. The main advantage of this approach is that it does not involve any Monte Carlo radiative transfer calculation, and hence generates no Monte Carlo noise. We generated a corresponding dust-free image for each observer and each broadband filter.

\subsubsection{Numerical parameter settings}

Apart from the definition of the sources, the dust medium, the instruments, and the probes, each SKIRT simulation requires a number of numerical parameters related to the different wavelength grids used internally or to the different Monte Carlo optimisation techniques. There was no reason to deviate from the default values, which are optimised to assure the best performance \citep[see also][]{2022MNRAS.516.3728T}.

A crucial parameter that directly affects the quality of the output data and the simulation run time is the number of launched photon packets. The optimal value was determined empirically using a limited number of galaxies of different stellar masses and sizes. We re-simulated the same galaxies with increasing number of photon packets until we found convergence on a pixel by pixel basis. Convergence was defined using the reliability statistics introduced in SKIRT by \citet{2018ApJ...861...80C, 2020A&C....3100381C}, based on work in the field of nuclear particle transport simulations \citep{MCNP2003}. In order to obtain convergence in the optical {\it{r}}-band out to at least twice the half-mass radius of the galaxies, we found that at least $10^9$ photon packets are required. For the FUV- and NUV-bands, the corresponding convergence area is typically smaller due to the lower intrinsic emissivity and the larger dust attenuation. We fixed the number of photon packets to $10^9$ for all galaxies. For a typical TNG50 galaxy the corresponding SKIRT run time on a dedicated 64-core machine was 3 to 4 hours, though there was a large variety depending on the number of stellar particles and the number of grid cells in the hierarchical octree grid. 


\section{The TNG50-SKIRT Atlas}
\label{ImageAtlas.sec}

\subsection{Atlas characteristics and availability}

For each of the 1154 TNG50 galaxies in our selection and for each of the 5 observing positions, the data set consists of dust-obscured and dust-free images in 18 broadbands, and 4 physical parameter maps. In total, this adds up to exactly 200 images per galaxy, or 230\,800 images in total. Each image or map is stored as an individual 1600 $\times$ 1600 pixel FITS file. The data and user guidelines on how to easily access and use them are available on the SKIRT website.\footnote{\url{https://skirt.ugent.be}} More details on the TSA characteristics are listed in Table~{\ref{AtlasChar.tab}}. 

The broadband images have units of MJy~sr$^{-1}$ and the parameter maps have natural units (${\text{M}}_\odot~{\text{pc}}^{-2}$ for the stellar and dust surface density maps, dimensionless units for the mean stellar metallicity maps, and Gyr for the mean stellar age maps). We note that the images are not convolved with a PSF and are noise-free, except for the presence of the Monte Carlo noise in the dust-obscured images. Users who wish to generate synthetic images matching a particular instrument or survey can convolve the images with the appropriate PSF and add background noise \citep[see e.g.][]{2019MNRAS.483.4140R, 2022MNRAS.511.2544D, 2023MNRAS.519.4920G}.

\subsection{Example galaxies}

\begin{figure*}
\includegraphics[width=\textwidth]{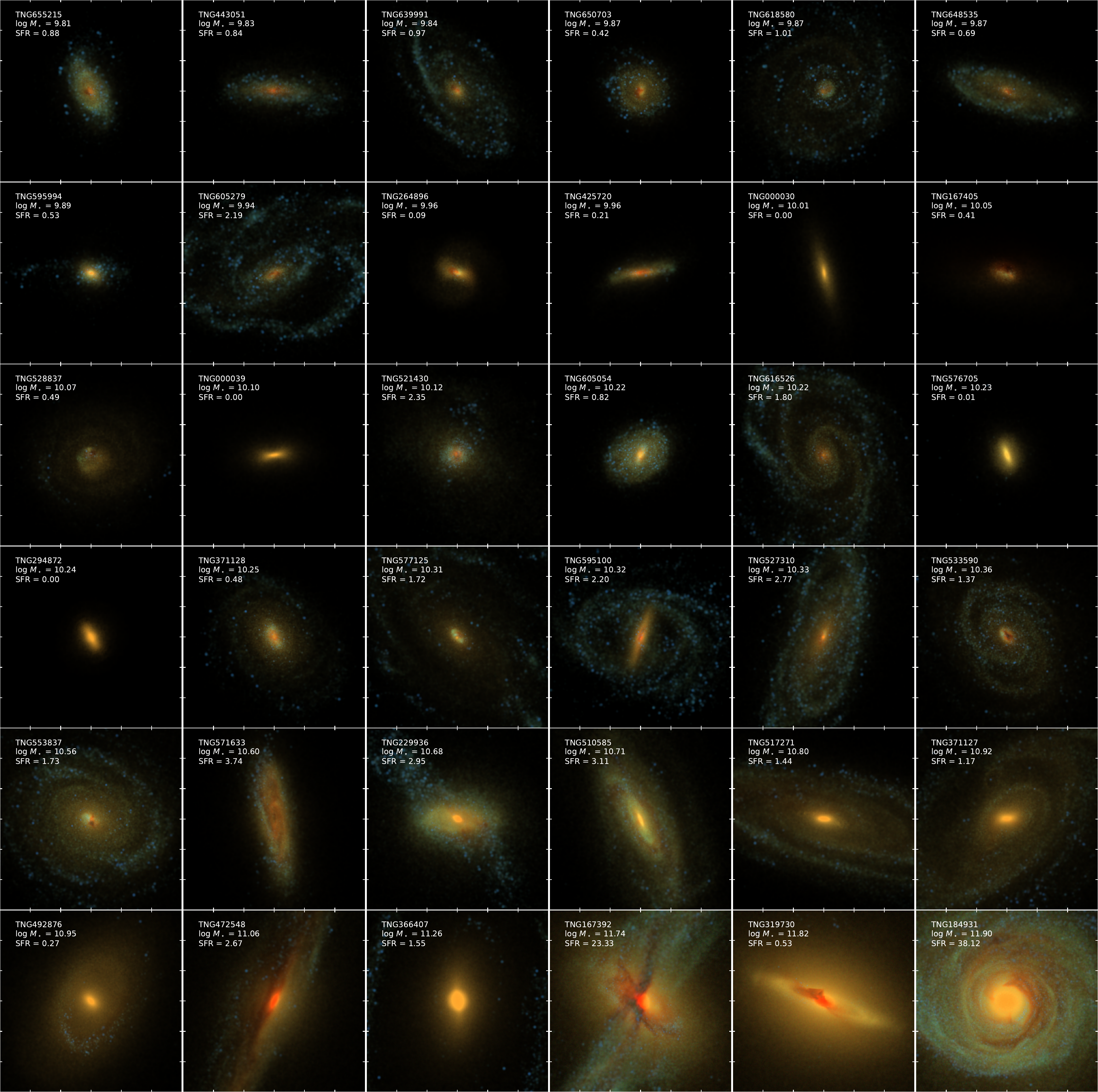}
\caption{Three-colour images for 36 randomly selected galaxies from our sample, all shown from observer position O1. The images are cutouts with a field-of-view of 30~kpc, and combine the LSST {\textit{u}}-, {\textit{g}}-, and {\textit{z}}-band images according to the methodology presented by \citet{2004PASP..116..133L}. The galaxies are sorted according to increasing stellar mass.}
\label{rgb.fig}
\end{figure*}

Fig.~{\ref{rgb.fig}} shows RGB images for 36 randomly selected galaxies, using the LSST {\textit{u}}-, {\textit{g}}-, and {\textit{z}}-band images for observer position O1. The galaxies are sorted by increasing stellar mass. This gallery illustrates the wide variety in optical morphology in the galaxy population: some galaxies are regular and almost featureless, others have a clear bulge-disc structure, and some have a disturbed morphology. In a number of galaxies, the effects of dust attenuation are clearly visible. Also the optical colours and SFRs of the galaxies vary widely, both globally and on spatially resolved scales.

\begin{figure*}
\includegraphics[width=\textwidth]{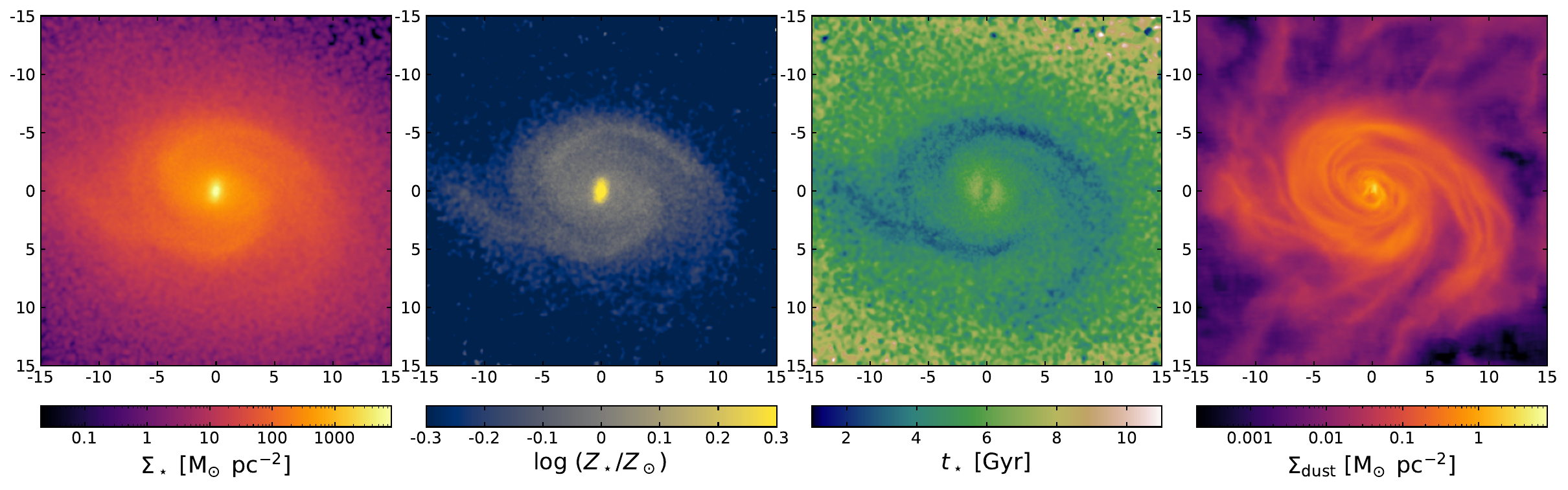}\\[-0.5em]
\includegraphics[width=\textwidth]{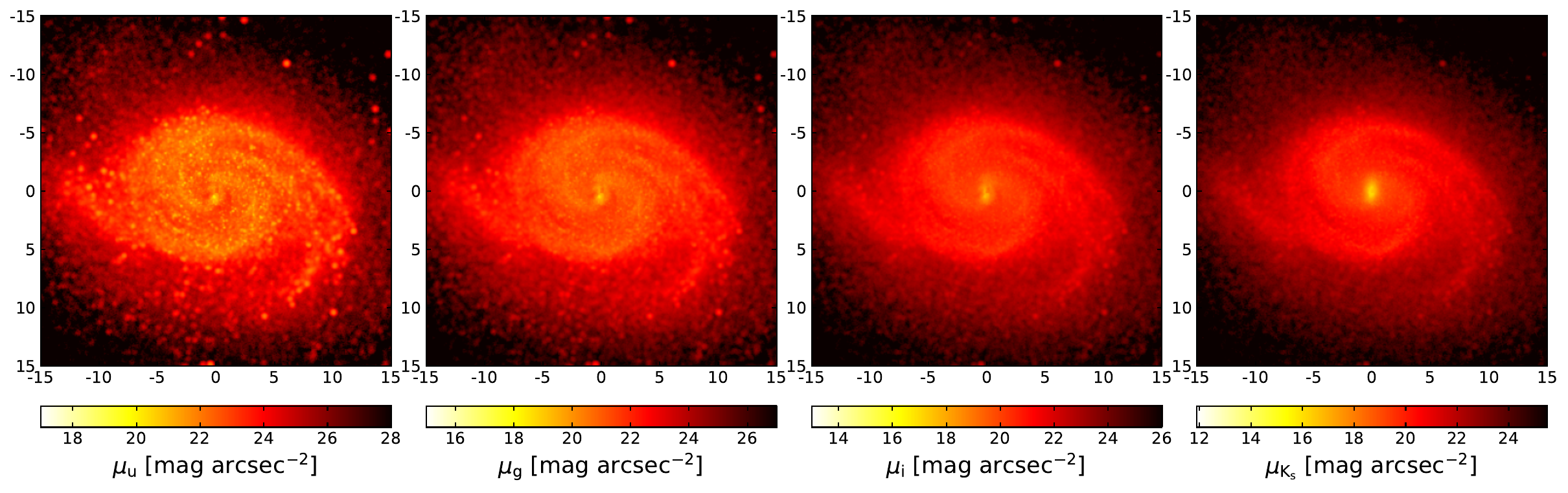}
\caption{Representative images and physical parameter maps for the spiral galaxy TNG000008 from observer position O2 ($i=47.6^\circ$). Top row: stellar mass surface density, mean stellar metallicity, mean stellar age, and dust mass surface density. Bottom row: {\textit{u}}-,  {\textit{g}}-,  {\textit{i}}-, and  ${\textit{K}}_{\text{s}}$-band images. All images and parameter maps zoom into the central $30\times30$~kpc region.}
\label{TNG000008_O2.fig}
\end{figure*}

\begin{figure*}
\includegraphics[width=\textwidth]{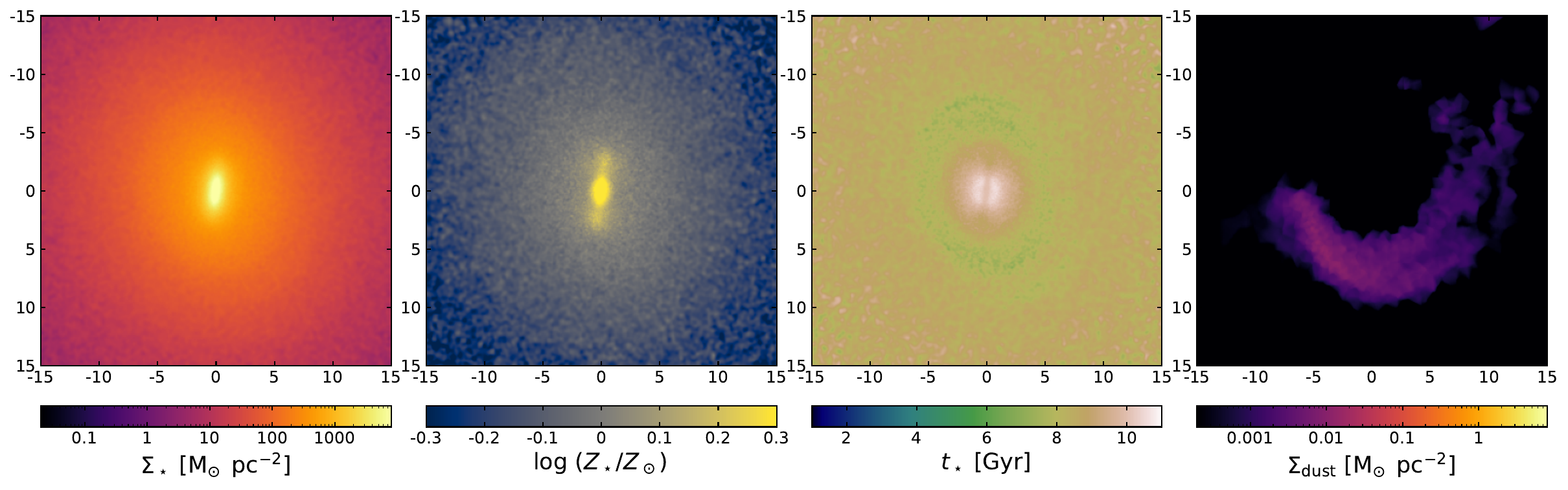}\\[-0.5em]
\includegraphics[width=\textwidth]{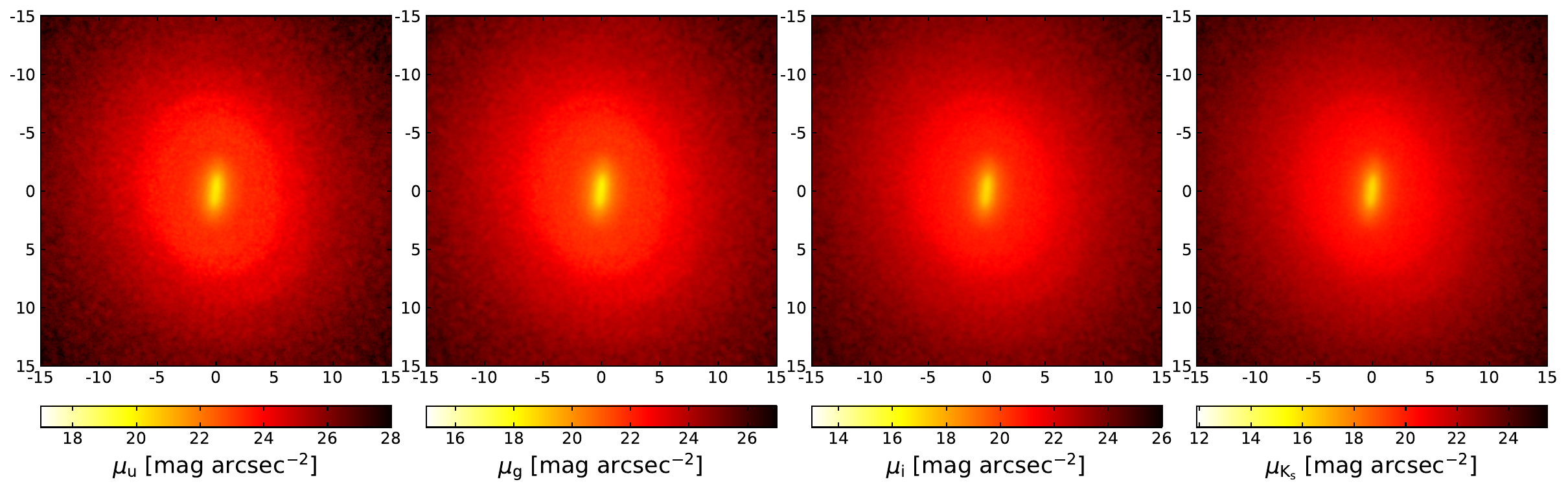}
\caption{Same as Fig.~{\ref{TNG000008_O2.fig}}, but for the massive early-type galaxy TNG096764 observed from observer position O1 ($i=35.9^\circ$). }
\label{TNG096764_O1.fig}
\end{figure*}

Figs.~{\ref{TNG000008_O2.fig}} and {\ref{TNG096764_O1.fig}} show a more detailed view on two handpicked galaxies from the data set. Fig.~{\ref{TNG000008_O2.fig}} shows TNG000008, a spiral galaxy with stellar mass $M_\star = 3.73\times10^{10}~{\text{M}}_\odot$ and ${\text{SFR}} = 3.98~{\text{M}}_\odot~{\text{yr}}^{-1}$, here seen at an intermediate inclination. The stellar mass distribution is characterised by a small but dense central component of metal-rich stars, whereas the disc is on average less metal-rich. The mass-averaged age of the stellar population is around 7~Gyr for the central bulge component and only about 3~Gyr for the populations in the spiral arms. The dust distribution is strongly concentrated in the central regions and the spiral arms and has a filamentary morphology. The broadband images show the signature as expected based on the physical parameter maps. The {\em{u}}-band image is dominated by the young stars and has a very clumpy appearance. It also shows clear signs of dust attenuation at the locations of the most prominent dust mass surface density enhancements. Moving to longer wavelengths, the images become less clumpy, the apparent signatures of attenuation gradually disappear, and the bulge gradually becomes more prominent. 

Fig.~{\ref{TNG096764_O1.fig}} shows a very different galaxy, TNG096764, a massive early-type galaxy with $M_\star = 1.05\times10^{11}~{\text{M}}_\odot$ and no ongoing star formation. The stellar mass surface density map shows a smooth distribution without obvious substructure, except a small bar in the central regions (viewed from this single observer's position, the central feature could be either a bar or an edge-on disc, but the combination of the different viewing angles shows that it is a bar rather than a disc). The stellar population shows a strong metallicity gradient, and the bar seen in the stellar mass surface density map stands out by its high metallicity. The same structure is also seen in the age map, where the feature stands out due to a slightly younger mean stellar age. On average, the stellar populations are relatively old and the age gradients modest. Furthermore, the galaxy displays a ring-like structure with a radius of about 5~kpc that is slightly younger ($\sim$6~Gyr) than the overall stellar population ($\sim$9~Gyr). Also interesting is the dust mass surface density map, which shows a conspicuous, albeit very low surface density, spiral-arm-like feature. Since there is no ongoing star formation and hardly any dust attenuation, the morphology of all broadband images is similar. It is very smooth without much substructure, except the inner bar and a faint shell-like structure in the bottom right quadrant. The {\it{u}}-band image of this galaxy is smoother than the {\it{u}}-band image of the spiral galaxy shown in Fig.~{\ref{TNG000008_O2.fig}}. 


\section{The {\textit{UVJ}} diagram}
\label{UVJ.sec}

\subsection{Comparison of TNG50, TNG100, and EAGLE}

\begin{figure*}
\centering
\includegraphics[width=\textwidth]{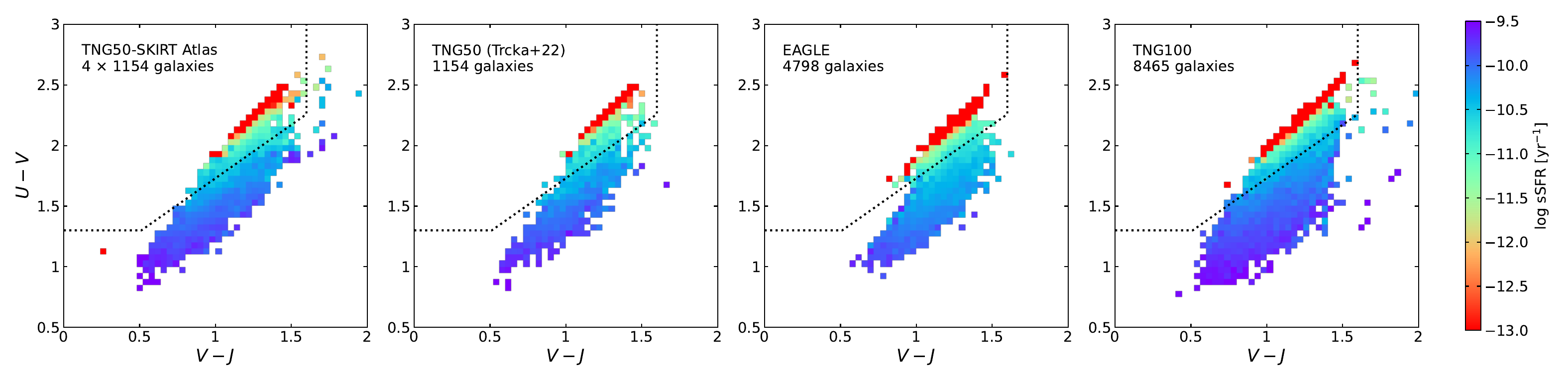}
\caption{The {\textit{UVJ}} plane of the TNG50, EAGLE, and TNG100 cosmological hydrodynamical simulations. Each panel contains only galaxies at $z=0$ within the same stellar mass range ($10^{9.8}~{\text{M}}_\odot < M_\star < 10^{12}~{\text{M}}_\odot$). All colours are calculated from SKIRT-generated fluxes or images that take dust attenuation into account. The colour scale represents the median sSFR of all galaxies with {\textit{UVJ}} colours within each pixel. The dotted line in each panel indicates the separation between the quiescent and star-forming galaxy populations and is taken from \citet{2019MNRAS.485.4817D}.}
\label{UVJ-comparison.fig}
\end{figure*}

\begin{figure*}
\includegraphics[width=\textwidth]{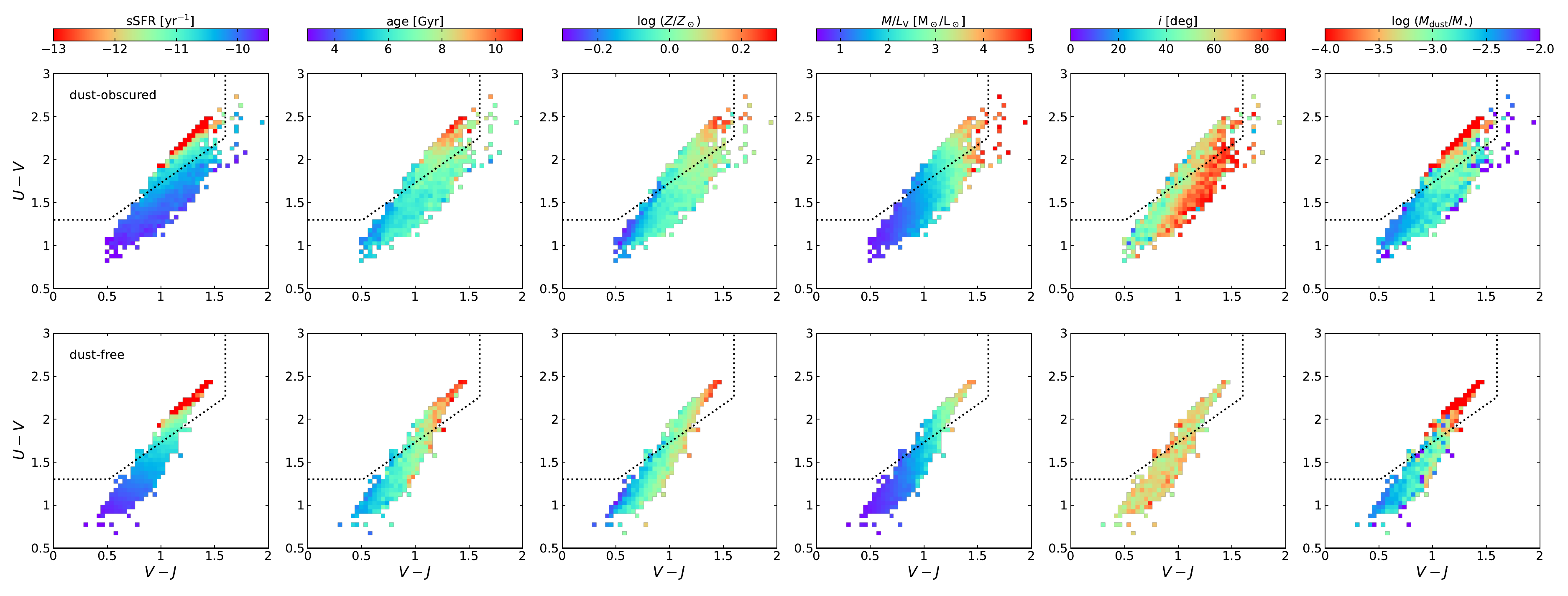}
\caption{Physical properties in the {\textit{UVJ}} plane, with (top row) and without (bottom row) dust attenuation. The different columns correspond to different physical parameters, indicated at the top of the column. The colour scale represents the median value of the physical parameter within each bin in {\textit{UVJ}} colours.}
\label{UVJ-global.fig}
\end{figure*}

The {\textit{UVJ}} diagram was first presented by \citet{2007ApJ...655...51W} and \citet{2009ApJ...691.1879W} as a way to distinguish between quiescent and star-forming galaxies. A single optical colour is not enough to separate these two classes as different combinations of dust attenuation and SFR can give rise to similar optical colours. However, the combination of an optical and an optical--NIR colour proves successful at making that distinction. The {\textit{UVJ}} method has been widely applied to select quiescent galaxies from observed galaxy samples \citep{2013ApJ...770L..39W, 2014ApJ...783L..14S, 2014ApJ...791...52B, 2015ApJ...803...26P, 2018ApJ...854...30P, 2018ApJ...858..100F, 2022ApJ...933...30T, 2023ApJ...947...20V} and from cosmological hydrodynamical simulations \citep{2017MNRAS.471.1671D, 2017MNRAS.470..771T, 2019MNRAS.485.4817D, 2022ApJ...929...94A, 2023arXiv231003083K}. 

The first panel of Fig.~{\ref{UVJ-comparison.fig}} shows the {\textit{UVJ}} diagram for the galaxies in our TSA. We calculated the rest-frame {\textit{U}}-, {\textit{V}}-, and {\textit{J}}-magnitudes by integrating the surface brightness of our images over the entire field-of-view. The colour scale in the {\textit{UVJ}} diagram represents the sSFR, which is taken directly from the TNG50 database. The other panels show equivalent {\textit{UVJ}} diagrams based on other simulated galaxy data sets, all at $z=0$ and within the same stellar mass range we used. The second panel contains the same set of galaxies, but used the `random' viewing position flux values generated by \citet{2022MNRAS.516.3728T}. The third panel is based on SKIRT-generated fluxes presented by \citet{2018ApJS..234...20C} for the flagship EAGLE simulation. The final panel corresponds to the TNG100 simulation, for which \citet{Gebek2024} generated broadband fluxes using the same methodology as applied by \citet{2022MNRAS.516.3728T} for the TNG50 simulation.

In all panels, the data show a clear correlation between the {\textit{UVJ}} colours and the sSFR. The dotted line, taken from \citet{2019MNRAS.485.4817D}, separates the quiescent galaxy population from the star-forming galaxies. For all four cases, the separation line corresponds roughly to a fixed ${\text{sSFR}} \sim 10^{-10.7}~{\text{yr}}^{-1}$. The general agreement between the different simulations is a good sanity check for the SKIRT calculations executed in this work. 

The differences between the different panels are at least as interesting. The main difference is the coverage of the {\textit{UVJ}} diagram, which can primarily be attributed to the sample size. The TNG100 fluxes generated by \citet{Gebek2024} reach lower ${\textit{U}}-{\textit{V}}$ colours for a fixed ${\textit{V}}-{\textit{J}}$ colour than the other diagrams. The galaxies populating these blue regions at the bottom of the {\textit{UVJ}} diagram are among the most actively star-forming galaxies. A remarkable difference is that the distribution of star-forming galaxies in our new TNG50 {\textit{UVJ}} diagram extends towards very red colours in the diagonal direction, beyond the vertical separation line at ${\textit{V}}-{\textit{J}}=1.6$, whereas it ends rather abruptly before this line in the other three panels (apart from some scattered galaxies in the TNG100 diagram). The galaxies populating this part of the {\textit{UVJ}} diagram are very actively star-forming with ${\text{sSFR}} > 10^{-10}~{\text{yr}}^{-1}$. At the same time they have very red colours, which implies that they must be heavily attenuated. It is, at first sight, surprising that these galaxies are lacking in the {\textit{UVJ}} diagram based on the fluxes calculated by \citet{2022MNRAS.516.3728T} since the galaxy samples used for the two TNG50 diagrams are exactly the same. The reason is that we consider four different random viewing positions for each galaxy, compared to a single random flux for the \citet{2022MNRAS.516.3728T} catalogue. As one can visually infer from Fig.~{\ref{rgb.fig}}, the level of dust attenuation in the TNG50 galaxies can differ substantially depending on the viewing angle, which can move galaxies across the {\textit{UVJ}} plane for different observer's positions. These red, star-forming galaxies are also missing from the EAGLE and TNG100 {\textit{UVJ}} diagrams, and this difference cannot be due to the different sample size. In these cases, the reason is probably the lower spatial resolution of the EAGLE and TNG100 simulations compared to the TNG50 simulation, which causes a puffier and more diffuse dust distribution, and therefore a shallower relation between attenuation and inclination \citep[e.g.][]{2017MNRAS.470..771T}.

A final note we observe that in our new TNG50 {\textit{UVJ}} diagram, and to a lesser degree also in the TNG100 diagram, we still note a clear anti-correlation between the ${\textit{U}}-{\textit{V}}$ colour and the sSFR for the galaxies beyond the vertical separation line at ${\textit{V}}-{\textit{J}}=1.6$. More specifically, the galaxies with the reddest ${\textit{U}}-{\textit{V}}$ colours are quiescent with ${\text{sSFR}} < 10^{-11}~{\text{yr}}^{-1}$, in spite of lying in the star-forming region of the {\textit{UVJ}} diagram. One could therefore question the shape of the quiescent galaxies region, as also raised by \citet{2019ApJ...874...17B}.

\subsection{Dust attenuation and physical properties in the {\textit{UVJ}} diagram}

While the {\textit{UVJ}} diagram is widely used to discriminate between star-forming and quiescent galaxies, its origin and reliability is being investigated in more detail \citep{2019ApJ...874...17B, 2019ApJ...880L...9L, 2019A&A...631A.156D, 2020ApJ...888...77W, 2022ApJ...939...29N, 2023ApJ...943..166A}. More specifically, it has been investigated to which degree the {\textit{UVJ}} colours can be used to constrain other physical properties of galaxies, and how dust attenuation affects the {\textit{UVJ}} colours of galaxies. The general effect of dust attenuation is that galaxies move upwards in the {\textit{UVJ}} diagram, roughly parallel to the diagonal separation line. The exact direction depends on the shape of the attenuation curve, which can vary significantly among galaxies \citep[e.g.][]{2016ApJ...827...20S, 2017ApJ...837..170L, 2018ApJ...859...11S, 2018ApJ...869...70N, 2022MNRAS.511..765Q, 2023MNRAS.524.4128Z}. The attenuation in our models is calculated in full 3D and takes into account absorption and scattering, which can sometimes lead to counterintuitive effects \citep{1992ApJ...393..611W, 1994ApJ...432..114B}. Moreover, our models have more complex, and hopefully more realistic, star formation histories than the parametric models often assumed.

In the top row panels of Fig.~{\ref{UVJ-global.fig}}, we show the {\textit{UVJ}} diagram based on the dust-obscured images for the galaxies in our sample colour-coded by six different physical parameters. The bottom row contains the same information, but now for {\textit{UVJ}} colours based on the dust-free images. The correlation between {\textit{UVJ}} colours and sSFR (see also Fig.~{\ref{UVJ-comparison.fig}}) is obvious for both the dust-obscured and dust-free diagrams, but it is not the only systematic trend. Also mean stellar age, mean metallicity, {\textit{V}}-band mass-to-light ratio, and dust-to-stellar mass ratio show a clear trend with the {\textit{UVJ}} colours. These results are well in agreement with the results obtained by  \citet{2019ApJ...880L...9L}. By means of a Bayesian inference method applied to synthetic SED models, they demonstrated that the mass-to-light ratio is well constrained by {\textit{UVJ}} colours alone, whereas the trends with age and metallicity are induced by galaxy scaling relations. 

\begin{figure}
\includegraphics[width=\columnwidth]{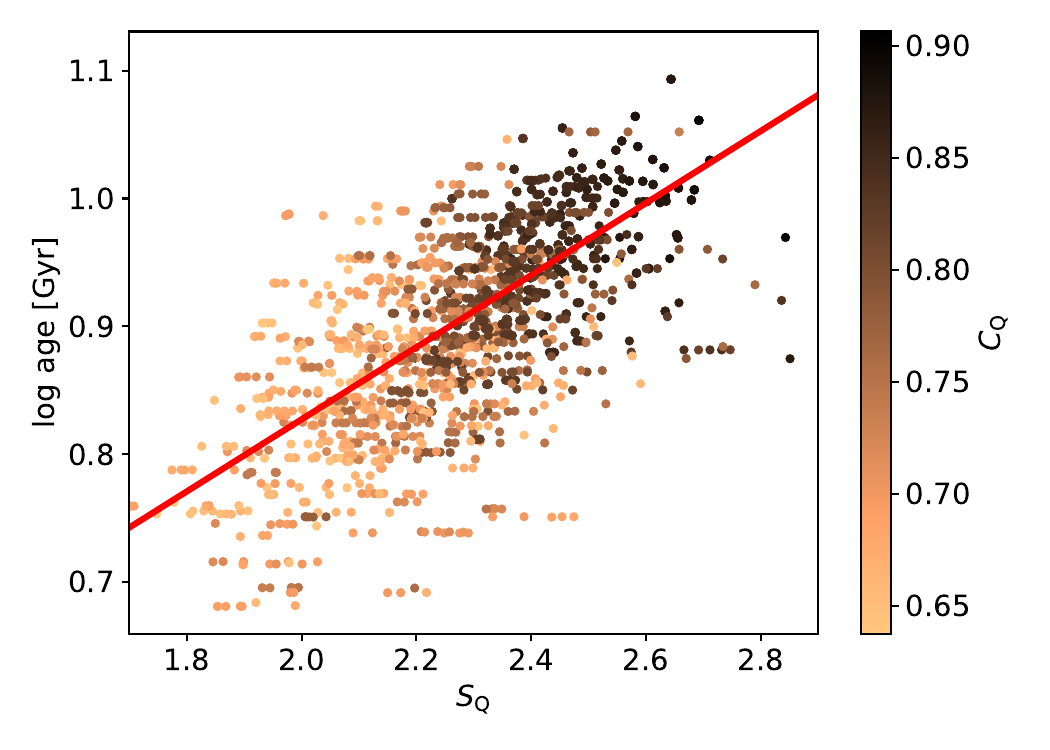}
\caption{Age--colour relation for quiescent galaxies in the TSA. The quiescent galaxy population is defined by means of the \citet{2019MNRAS.485.4817D} separation line in the {\textit{UVJ}} diagram.}
\label{age-colour.fig}
\end{figure}

Looking at the quiescent galaxies region of the second panel on the top row, one can see that the mean stellar ages of the galaxies systematically increase when moving upward in the diagonal direction. To investigate this in more detail, we used the rotated coordinate system on the {\textit{UVJ}} diagram introduced by \citet{2018ApJ...858..100F}. In this system, the rotated axes $S_{\!\text{Q}}$ and $C_{\text{Q}}$, defined as
\begin{gather}
S_{\!\text{Q}} = 0.75\,(V-J) + 0.66\,(U-V),
\\
C_{\text{Q}} = -0.66\,(V-J) + 0.75\,(U-V),
\end{gather}
run parallel and perpendicular to the boundary of the quiescent box, respectively. In Fig.~{\ref{age-colour.fig}} we show the relation between the colour index parallel to the separation line and the logarithm of the mean stellar age of each galaxy in the quiescent region. We recovered a reasonably strong linear correlation ($R^2 = 0.44$), which means that the ages of quiescent galaxies can in principle be estimated from {\textit{UVJ}} colours alone. This supports the conclusions by \citet{2019ApJ...874...17B}, who derived a similar linear relation between the logarithmic median stellar age and the $S_{\!\text{Q}}$ index. 

Comparing the panels on the top row with the corresponding ones on the bottom row, a number of interesting effects of dust attenuation can be discerned. One can immediately note the difference in coverage of the {\textit{UVJ}} plane, which is most obvious for the star-forming galaxy population. The dust-free star-forming galaxies predominantly occupy a region with $0.5\lesssim {\textit{V}}-{\textit{J}} \lesssim 1.1$ which moves into the quiescent galaxies region as soon as ${\textit{U}}-{\textit{V}} \gtrsim 1.7$. The dust-obscured star-forming galaxies, on the other hand, occupy a much more extended region that runs parallel to the separation line up to ${\textit{V}}-{\textit{J}} > 1.5$. The galaxies in this new territory have high sSFR values, and so they moved from the bottom left corner in the dust-free {\textit{UVJ}} diagram. Interestingly, a large number of these galaxies are oriented almost edge-on, in line with the findings by \citet{2012ApJ...748L..27P}. This segregation by inclination is clearly visible when comparing the panels in the fifth column. While the galaxies in the bottom panel have a uniform distribution in inclination with a median value of about 60$^\circ$ in every pixel of the parameter space, the distribution is clearly separated in the top panel, in particular for the star-forming galaxies: low-inclination galaxies remain at the blue side of the diagram, whereas high-inclination galaxies move towards the red side. At the extreme end of the star-forming distribution we find edge-on galaxies with high sSFR values and very high dust-to-stellar-mass ratios.

There is also a difference in the dust-free and the dust-obscured {\textit{UVJ}} diagram for the quiescent galaxy population. In the dust-free diagram it forms a well-defined sequence out to ${\textit{V}}-{\textit{J}} \approx 1.5$, populated exclusively by galaxies with very low sSFR values, old ages, high metallicities, large mass-to-light ratios, and very small dust-to-stellar-mass ratios. When dust attenuation is turned on, however, the population extends to redder colours, but most importantly, the region in the {\textit{UVJ}} space below the original tight sequence is populated, across the boundary line into the star-forming galaxies regimes. The galaxies in this new region are, again, oriented almost edge-on and heavily attenuated. We can quantify the nature of these additional galaxies by calculating some global statistics of the two different populations (as defined by the dotted line region in the plots) with and without dust attenuation. In the dust-free case, the quiescent galaxy region contains just 36\% of the total stellar mass budget, and the mean sSFR and stellar age are $\langle {\text{sSFR}} \rangle = 10^{-11.52}~{\text{yr}}^{-1}$ and $\langle t \rangle = 8.84$~Gyr. In the dust-obscured case, the pollution by star-forming galaxies increases the fraction of the stellar mass budget of the quiescent galaxy region to 48\%. The mean sSFR increases by 0.4~dex to $\langle {\text{sSFR}} \rangle = 10^{-11.12}~{\text{yr}}^{-1}$, and the mean stellar age decreases to $\langle t \rangle = 8.55$~Gyr. In summary: dust attenuation pollutes the quiescent galaxy region with younger and more actively star-forming highly inclined galaxies.


\section{Discussion and conclusion}
\label{Conclusions.sec}

\subsection{Summary}

The ambition of this work was to generate, present, and publicly release a synthetic UV--NIR broadband image atlas for a complete stellar-mass selected sample of 1154 galaxies extracted from the $z=0$ snapshot of the TNG50 cosmological simulation \citep{2019MNRAS.490.3196P, 2019MNRAS.490.3234N}. The images were generated with the SKIRT radiative transfer code \citep{2015A&C.....9...20C, 2020A&C....3100381C} and account for different stellar populations and absorption and scattering by interstellar dust in a realistic 3D setting. 

For each galaxy, we generated a suite of 100~pc resolution images in 18 broadband filters, for five different observer positions. In addition to the dust-obscured images, we also released synthetic images without dust attenuation, and stellar mass surface density, mean stellar age, mean stellar metallicity, and dust mass surface density maps. While other teams have already released synthetic image datasets for cosmological hydrodynamical simulations \citep[e.g.][]{2015MNRAS.447.2753T, 2015MNRAS.452.2879T, 2019MNRAS.483.4140R, 2021MNRAS.506.5703K, 2022MNRAS.512.2728C, 2023MNRAS.519.4920G}, we argue that the present atlas is unique in its kind due the realism of both the underlying simulation and the radiative transfer treatment, the large sample size, the high spatial resolution, the number of filters, and the combination of matching images and physical parameter maps. 

As a sanity check and a first application of our image atlas, we investigated the {\textit{UVJ}} diagram. Comparing our {\textit{UVJ}} diagram with the one based on the fluxes generated by \citet{2022MNRAS.516.3728T} for the same TNG50 galaxy sample, and with the EAGLE and TNG100 {\textit{UVJ}} diagrams we found excellent agreement in terms of the relation between sSFR and {\textit{UVJ}} colours. The diagrams also show some interesting differences, mainly in the coverage of the {\textit{UVJ}} diagram. In particular, the distribution of star-forming galaxies in our new TNG50 UVJ diagram extends towards very red colours in the diagonal direction, whereas the other diagrams lack these very actively star-forming and heavily obscured galaxies. These differences can be interpreted as a result of different sample size and the difference in spatial resolution. 

We also investigated the trends of galaxy physical parameters in the {\textit{UVJ}} diagram. We found that, apart from the strong correlation with sSFR, the {\textit{UVJ}} colours also show systematic trends with mean stellar age, mean stellar metallicity, {\textit{V}}-band mass-to-light ratio, and dust-to-stellar-mass ratio. We found a reasonably strong positive correlation between the mean stellar age and the {\textit{UVJ}} colours for the quiescent galaxy population, which suggests that the ages of quiescent galaxies can be well constrained by {\textit{UVJ}} colours alone \citep{2019ApJ...874...17B}. Finally, we investigated the effect of dust attenuation on the distribution of the galaxy population in the {\textit{UVJ}} diagram. As expected, dust attenuation spreads the galaxy population towards redder colours parallel to the separation line between the quiescent and star-forming galaxy populations. In the dust-free {\textit{UVJ}} diagram the quiescent galaxy population forms a well-defined sequence populated exclusively by galaxies with very low sSFR values and old ages, but dust attenuation pollutes this quiescent galaxy region with younger and more actively star-forming galaxies. Dust attenuation generates a clear separation in inclination of the star-forming galaxies: low-inclination galaxies remain at the blue side of the diagram, whereas high-inclination galaxies move towards the red side. The reddest star-forming galaxies are edge-on, dusty galaxies with high sSFR values.

\subsection{Possible applications}

The prime goal of this paper was to present and publicly release the TSA. We demonstrated its usefulness by investigating the {\textit{UVJ}} diagram. We hope that this image atlas can be used for many more applications.

A prime application is the connection between the morphology of galaxies, their fundamental physical properties, and the environment in which they reside \citep[e.g.][]{2003ApJS..147....1C, 2014ARA&A..52..291C, 2009ARA&A..47..159B, 2021gamo.book.....H}. The availability of multi-band dust-obscured and dust-free images allows for a systematic investigation of the wavelength dependence of galaxy morphology \citep{2009ApJ...703.1569M, 2012MNRAS.421.1007K, 2014MNRAS.441.1340V, 2020A&A...641A.119B, 2023A&A...673A..63N, 2023ApJ...957...46M} and the effects of dust attenuation on photometric and morphological parameters \citep{2010MNRAS.403.2053G, 2013A&A...553A..80P, 2013A&A...557A.137P, 2023MNRAS.524.4729S}. 

In a companion paper to this release paper \citep{Baes2024b} we used the TSA to investigate the wavelength dependence of the effective radii of galaxies. In the near future we plan to employ single or multiple-component S\'ersic fitting and non-parametric morphological indices to quantify morphology. All global physical properties (also including dark matter properties), the intrinsic particle and cell data, and the entire history for all galaxies can be readily accessed through the TNG public database \citep{2019ComAC...6....2N}, which allows for a thorough investigation on what drives galaxy morphology. 

In the past few years, several well-known global galaxy scaling relations have been investigated on local, sub-kpc scales as well. Examples include the local dust scaling relations \citep{2014A&A...567A..71V, 2022A&A...668A.130C}, the spatially resolved star-forming main sequence \citep{2016ApJ...821L..26C, 2017ApJ...851L..24H, 2020MNRAS.496.4606M, 2022A&A...663A..61P, 2022ApJ...935...98A, 2022MNRAS.510.3622B}, the resolved stellar mass gas metallicity relation \citep{2012ApJ...756L..31R, 2016MNRAS.463.2513B, 2018ApJ...868...89G, 2023MNRAS.520.4301B, 2023MNRAS.519.1149B}, or the resolved stellar mass stellar metallicity relation \citep{2014ApJ...791L..16G, 2020MNRAS.491.3562Z, 2021MNRAS.508.4844N, 2022MNRAS.512.1415Z, 2023A&A...673A.147P}. Our image and physical parameter map database can be used to investigate the universality of these local scaling relations and to determine the physical scales at which they potentially break down. 

We plan to address several of the questions raised above. However, we also publicly release these data and warmly invite the community to use them in any way they see fit.

\subsection{Caveats and future work}

While we believe that the current image atlas is sufficiently rich and realistic to allow for a range of applications, we are aware that it also has its caveats and limitations. A first important aspect is that our data are built on simulated galaxies extracted from the TNG50 cosmological hydrodynamical simulation. While this simulation is generally considered as one of the most powerful large-volume simulations, it comes with its own caveats and limitations. One of them is that the TNG model was designed and calibrated at the resolution of the original Illustris simulation, while the TNG50 simulation has a roughly 16 times better mass resolution. This improved resolution results in somewhat larger galaxy masses and SFRs \citep{2018MNRAS.473.4077P, 2018MNRAS.475..648P, 2019MNRAS.490.3196P, 2019MNRAS.485.4817D, 2022MNRAS.516.3728T}.

A second limitation of our atlas is that it is limited to the UV--NIR range, that is, the range where stellar emission dominates the SED. This limitation is not inherent to SKIRT code: the code has been used to generate synthetic SEDs and images for galaxies that cover the entire UV-mm wavelength range. The main reason is computational: to simulate the UV--NIR range only dust attenuation and no dust emission is required. SKIRT simulations with dust emission are computationally much more demanding (up to an order of magnitude, depending on the details of the simulation) compared to attenuation-only simulations. 

In previous post-processing work of cosmological hydrodynamical simulations, we encountered difficulties to reproduce the UV and MIR fluxes and colours of observed galaxies: we typically found insufficient UV attenuation and too much emission in the MIR \citep{2019MNRAS.484.4069B, 2020MNRAS.494.2823T, 2022MNRAS.516.3728T, 2021MNRAS.506.5703K, 2022MNRAS.512.2728C}. We argued that the MAPPINGS~III templates we use for the young stellar particles are at least partly responsible for these discrepancies. \citet{2023MNRAS.526.3871K} recently generated a new template library, called TODDLERS, to be used in SKIRT with the aim of addressing this problem. The first tests of this new library are very promising (Kapoor et al., in prep.). Our ambition is to rerun our image library with this new emission library in the near future, extending the range to mm wavelengths. This new emission library is expected to not affect the optical images, but will probably improve the UV images. We advise users to take this caveat into account when they use our data. 

Looking forward, we see this UV--NIR broadband image atlas as an intermediate step in an effort to generate increasingly more realistic synthetic data products for simulated galaxies. As discussed above, our next ambition is to extend this image database to the UV--mm wavelength range, incorporating the new TODDLERS library. The step beyond that could be a transition from broadband imaging to full spectral resolution. Such an effort could be similar to the works of \citet{2022MNRAS.514.2821B}, \citet{2022MNRAS.515..320N, 2023MNRAS.522.5479N}, and \citet{2023A&A...673A..23S}, but with a completely self-consistent treatment of dust attenuation and emission, and ideally covering the entire UV--mm wavelength range. Several intermediate steps towards realising this ambition have already been taken or are under development: the TODDLERS templates have full spectral resolution and contain a detailed treatment of nebular emission lines, SKIRT has full support for gas and stellar kinematics \citep{2020A&C....3100381C, 2023MNRAS.524..907B}, and we are currently working on a sub-grid model for the multi-phase ISM, inspired by \citet{2021ApJ...922...88O} and \citet{2021A&A...645A.133R, 2023A&A...679A.131R}.

Finally, applying the SKIRT radiative transfer post-processing recipe for different redshift snapshots would open up the possibility to directly investigate and test the cosmic evolution of galaxy morphology and many of the scaling relations mentioned above.


\begin{acknowledgements}
MB, NA, IK, and MM acknowledge financial support by the Flemish Fund for Scientific Research (FWO-Vlaanderen) through the research projects G030319N, G0G0420N, and G037822N.  AG and BVM acknowledge the support by FWO-Vlaanderen through the PhD Fellowship Grants 11H2121N and 1193222N, respectively. CT acknowledges the INAF grant 2022 LEMON. This project has received funding from the European Research Council (ERC) under the European Union's Horizon 2020 research and innovation program (grant agreement No.\ 683184).
\end{acknowledgements}


\bibliography{mybib.bib}

\end{document}